# Preproduction Deploys:
# Cloud-Native Integration Testing

Jeremy J. Carroll
Infrastructure
Coursera
Mountain View
jcarroll@coursera.org

Pankaj Anand
Infrastructure
Coursera
Mountain View
panand@coursera.org

David Guo
Infrastructure
Coursera
Mountain View
dguo@coursera.org

*Abstract*—The microservice architecture for cloud-based systems is extended to not only require each loosely coupled component to be independently deployable, but also to provide independent routing for each component. This supports canary deployments, green/blue deployments and roll-back. Both ad hoc and system integration test traffic can be directed to components before they are released to production traffic. Front-end code is included in this architecture by using server-side rendering of JS bundles. Environments for integration testing are created with preproduction deploys side by side with production deploys using appropriate levels of isolation. After a successful integration test run, preproduction components are known to work with production precisely as it is. For isolation, test traffic uses staging databases that are copied daily from the production databases, omitting sensitive data. Safety and security concerns are dealt with in a targeted fashion, not monolithically. This architecture scales well with organization size; is more effective for integration testing; and is better aligned with agile business practices than traditional approaches.

*Keywords—cloud computing, microservices, software architecture, software integration testing*

## I. Introduction

Coursera provides a platform for the delivery of cloud-based online education.

We use a Continuous Integration/Continuous Deployment [CI/CD] pipeline [1] updating one of the hundreds of loosely coupled components every few minutes. In 2018, as we increased the number of degrees being offered online via our platform [2], we decided to validate every change before it went live. We use a suite of Puppeteer [3] integration tests, to verify over 400 user facing tasks for each change, using browser automation.

This paper addresses the challenge of how to establish and maintain a scalable preproduction environment to perform this validation. Test traffic is routed through preproduction deploys [4] within the production cloud before these deploys are released to production traffic. The preproduction deploys are isolated from the production system, by using staging databases, that are copied daily from the production system. Use of the staging databases provides sufficient isolation, and by using this approach, you do not need a complete staging environment.

The technical foundations are: the separation of the deploy step from the release step [5]; and extending the microservice architecture [6], to not only require components to be independently deployable, but also to support multiple concurrent versions with independent traffic routing, on a per request basis. The routing information selecting microservice versions and the staging databases is attached as annotations to each request, as it passes through the system.

This forms a cloud-native development (CND) architecture: this paper presents integration testing within this architecture.

The reader wishing to add scalable integration testing during the deploy of their CI/CD pipeline can first read the overview (IV) and then skip to implementation considerations at the end (VIII), referring back to the more detailed sections as needed.

Combined with Agile methodologies [7], the CND architecture supports dynamic business responses such as Coursera's nimble response to the Covid-19 Pandemic, with a swift rework of the Coursera for Campus product to address the new needs of universities around the world as they had to adjust to remote teaching [8].

## II. Problems Addressed

Preproduction deploys are used for integration testing, ad hoc testing, troubleshooting and review. This paper will show that, compared with traditional development and staging environments, this architecture is:

- more effective, abolishing "Cannot reproduce" bugs.
- much easier to maintain since there is a unified approach to preproduction and production.
- cheaper in cloud resources since test traffic is mostly handled by elastic scaling in the production components.
- more flexible.
- more amenable to incremental continuous development, without bottlenecks from serializing changes into an integration test queue.

## III. Background

### A. Cloud-Native Microservices

Some technologies and approaches work better than others when developing cloud-based services. This gives rise to the notion of "cloud-native": characterized by [9], as "a distributed, elastic, and horizontally scalable system composed of services and operated on self-service platforms, while the services itself [sic.] are designed as self-contained deployment units according to cloud design patterns" (from [10]). The principal cloud design pattern is microservices: "designing software applications as suites of independently deployable services" [6]; "each microservice [is] an independent unit of development, deployment, operations, versioning, and scaling" [11].

At Coursera, we use microservices within a service mesh [12]. "A service mesh is an infrastructure layer devoted to managing, observing, visualizing, and controlling [micro]services to make their intercommunication safe and reliable," [13]. A specific feature of service meshes is that they allow multiple deployed versions of each microservice, with the version selected for any specific request determined either using rules or randomly.

The Cloud Native Computing Foundation suggests attributes of appropriate technology [14] as "resilient, manageable, and observable" in contrast to traditional approaches that may be "hard to scale; not fault-tolerant; not self-healing; inefficient due to poor utilisation of resources".

We use preproduction deploys within the production cloud so that these advantages of cloud-native approaches are also present during integration testing.

### B. Integration Environments

Every software development team needs a test environment to verify the code that they have written together. The goal is to understand what the code will do once it is in the production environment.

Prior to online services, the production and testing environments were physically separate. This separation was a fact of life in the 1970s and 1980s, before the business use of the Internet. Cloud software has inherited this so-called 'best practice', often justified by the need to keep production 'safe', without reflection as to the cost/benefit trade off, or the actual threats to the safety of production.

The inevitable differences in hardware and capacity led to "Works for Me"/"Cannot Reproduce" bugs, where the code behaved differently in the different environments, estimated at about 4% of all bugs, [15].

More recently developers and test engineers have used virtual machine environments *as similar to the production environment as possible* [1]. The large code footprint and number of dependencies of cloud systems makes maintaining such a VM environment very difficult. The different compute, storage, networking, latencies and load between the VM and production make the experiences significantly dissimilar.

An alternative approach is to use a release train [7], which at a given cadence clones the production environment in its entirety as a staging environment, with a new version of all the software prior to updating production. In practice, this clone is far from perfect. In addition, each developer needs to ensure that their code will work correctly with code being added to the release train by other developers in other teams. Otherwise, the integration testing should fail, and one or other pieces will be rejected from the monthly release (which may be entirely derailed).

In contrast, this paper advocates preproduction deploys for these functions, since "the *best effort simulation* of the production environment is … the production environment itself." [16].

### C. "Testing in Production"

A new paradigm, testing in production [16], [17], has been arising. While our approach is influenced by such work, there are important distinctions. One is that the testing in production paradigm uses production traffic to validate new software components deployed in the cloud, while we validate preproduction deploys in the cloud using test traffic, prior to any exposure to production traffic. A second is that we use staging databases to isolate test and preproduction usage from production usage within the production cloud.

### IV. OVERVIEW

As per [5], [16] we use the terms *deploy* and *release* (both as verbs and nouns) with precise meanings: to *deploy* some software is to build infrastructure (the *deploy*) in the cloud for one version of the software; to *release* some software is to use that *deploy* to service customer requests. Such a *deploy* is known as a *release*. With these terms we summarize the rest of the paper. Within the CND architecture:

- Various types of components can be independently deployed.

- Each component consists of multiple cloud resources, and a software codebase.

- Different components are loosely coupled.

- The resources within a component are tightly coupled.

- Multiple concurrent preproduction deploys can be made of each component.

- One deploy of each component is marked as the current production release.

- Routing software controls which deploy of each component is used on a per external request basis.

- Production requests are routed to the production release of each component.

- Test requests and integration tests are annotated with routing information specifying preproduction deploys of zero, one or more components.

- Test requests are routed to the specified preproduction deploys for the specific components, and to the production releases of everything else.

- Staging databases are copied from production databases at regular intervals, omitting sensitive data.

- Production requests are routed to production databases.

- Test requests are routed to the staging databases.

- Only preproduction deploys made from the main branch can be released for production traffic.

- Only preproduction deploys that pass the integration tests can be released.

## V. An Architecture for Preproduction

### A. Attributes of Cloud Native Applications and Components

Cloud native applications have characteristic signs, we combine those found in [9], [10], [14], emphasizing a few that are important in this paper … and then adding three more, not found in those references.

#### 1) From the literature

Cloud native applications are typically formed from a collection of *loosely coupled components* following a microservice architecture. These components normally isolate their state, may be deployed in containers, and may use a service mesh to abstract the interservice network communication, often based on REST principles. The deployment process for each component is largely *self-contained*.

The infrastructure used can be abstracted and managed as code rather than directly and is usually both immutable and disposable.

Both the design of each component and the infrastructural choices may support scalability and policy-driven elasticity: providing fault-resilience and self-healing. The utilization of resources is usually efficient.

Tools manage the deployed components, providing *observability and monitoring* through the use of centralized distributed logging and metrics.

The whole process is integrated into a *CI/CD pipeline* that deploys and releases each component.

#### 2) Our focus

To achieve cloud-native integration testing using the methods of this paper, in addition to the emphasized items above, you also need:

- Fast rollback for each component
- Ability to separate deploying a component from releasing that deploy to production traffic, see IV.
- Ability to route an individual request through a specific deployed version of each component.

The first, with good monitoring, enables speedy recovery from mistakes, which allows a thoughtful acceptance of risk. The second is a key enabler for fast rollback, since it allows two systems to be deployed, and one to be released to production traffic. The third enables gradual rollout in a blue-green deploy [18], and canary analysis [1] (see V.C concerning both of these).

We work within a Continuous Integration/Continuous Deployment that starts with a process of small incremental changes at every part of the development process. This process is not dissimilar to Continuous * [19]: the whole business operates in accord with agile principles.

### B. Types of Service Components

To facilitate continuous integration and deployment across our platform, we use the microservice principle of independent deployability for almost all our components, not just services. Some important component types are:

- Service mesh [12] internal microservices: typically presenting an HTTP/REST interface to other microservices within the mesh.
- Gateway microservice(s): the ingress to the service mesh receiving traffic from outside the mesh and servicing it by making requests inside the mesh.
- JavaScript Single Page Applications.
- Near-line stream processors of events in a message queue published to by microservices (at Coursera these sit inside the service mesh).
- Off-line processes (for **cron** based processing of the production data), for example workflows using Apache Airflow.
- The integration testing components.
- The deployment software.

Each component consists of multiple cloud resources that work in a tightly coupled fashion to deliver some part of the overall function. Different components are loosely coupled. Most components are elastic, and can be easily sized up or down, often in real-time, fully automatically.

For each component type we support the previously listed attributes, see V.A. As we gradually roll out the architecture, different components have different levels of automation supporting each of these attributes.

### C. Deploy And Release

A deployed component that has not been released is configured to receive no traffic other than test traffic that is explicitly routed to it. For microservices this configuration is made in the service mesh.

While in an emergency, or for rapid rollback, it is possible to immediately route 100% of the production traffic to a deployed component, hence fully releasing it, normally a more conservative approach is used. For example, we might start with a canary release, in which maybe 1% or 10% of the production traffic is routed to the component, for some minutes. The error rate and latency are monitored of the canary release during this period and compared with the previous production release. If there is a significant unexpected increase in either, then the release is aborted, and the previous release once again serves 100% of the traffic.

If the canary is successful, the process of blue-green deployment gradually increases the percentage of traffic going to the new release, with corresponding decreases going to the previous release. If the error rate or latency shows poor behavior then once again there is an opportunity to abort.

The benefit of using this approach is that even if other development processes have failed to catch errors before release, the amount of production traffic that is impacted is small.

Both canary releases and blue-green deployments are used widely in cloud native applications. Neither strictly require the flexibility that our CND architecture asks for, but given that flexibility these are easily implemented. In addition, the CND architecture allows for a comprehensive suite of integration tests

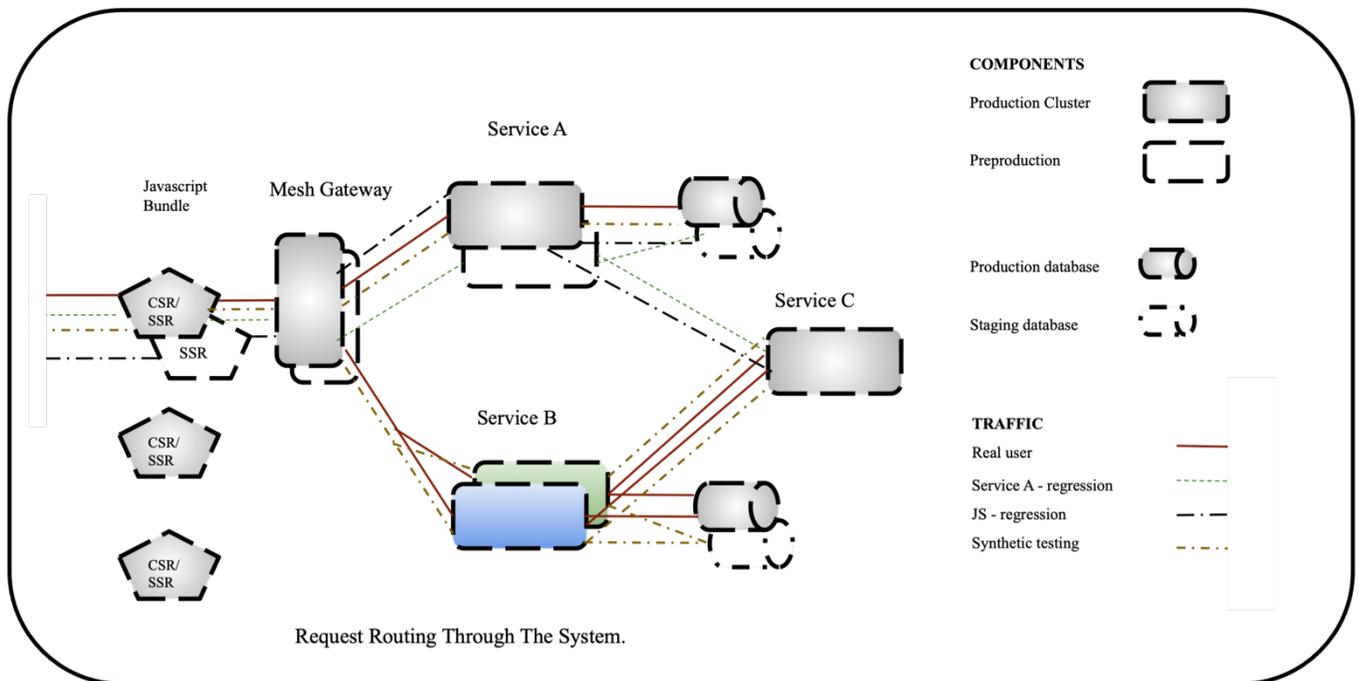

Request Routing Through The System.

to be run between deploy and release, to give increased confidence before any production traffic is directed to a new release.

This gives a more detailed upgrade process like:

- Deploy the main branch as a preproduction.
- Run integrations tests with the preproduction deploy.
- Verify the tests passed.
- Canary the new deploy with 1% or 10% of the production traffic.
- Verify the canary is successful.
- Gradually release all production traffic to the new deploy.
- Verify the release succeeded.
- Delete the old deployment (we usually wait some hours for this step, to allow for rapid rollback if problems with the new release are slow to exhibit).

*D. Request Routing*

The component update process relies on the ability to have multiple versions of the same component and to dynamically switch traffic both intentionally and randomly between them.

This routing is controlled by annotating external and internal traffic, e.g., with headers or cookies. Any external annotation is cryptographically signed, to ensure traffic routed to preproduction systems is legitimate test traffic.

In figure 1, there is a system consisting of multiple components. Each component may have multiple versions, which are illustrated overlapping with one another. Each component is shown as a single box in the diagram, which represents multiple tightly coupled resources working together.

- The service mesh gateway manages the request routing through the service mesh.
- Server-side rendering (SSR) [20] manages front-end routing.
- Html creation for client-side rendering (CSR) [20] selects a specific bundle version for each JavaScript single page application.
- The version of the mesh ingress microservice can be selected using its IP address.
- Routing information is included with each event for near-line processing (not shown), so that all request processing uses the selected versions of cloud microservices.
- The integration test framework seamlessly supports each of these routing methods, so that typically the developer takes no manual steps to control routing.

On the left, there are three of hundreds of different JavaScript bundles – that can be downloaded and run in a browser or run server side. A preproduction version is running with SSR in the cloud. Production traffic, and synthetic testing, is hitting the production version, as is test traffic for backend services.

TABLE I. DEPLOYS AND RELEASES AT COURSERA

| Component | Activity | Week of May 24th, 2021 | | | | |
|---|---|---|---|---|---|---|
| | | *Mon.* | *Tue.* | *Wed.* | *Thu.* | *Fri.* |
| Gateway microservice | deploy | 0 | 1 | 3 | 2 | 0 |
| | release | 0 | 1 | 1 | 2 | 0 |
| Cloud microservice | deploy | 23 | 40 | 17 | 23 | 10 |
| | release | 21 | 31 | 17 | 19 | 9 |
| JavaScript Application | deploy | 1243 | 843 | 802 | 899 | 261 |
| | release | 18 | 12 | 58 | 67 | 10 |

These frontends all send traffic to the backend via the production mesh gateway. A preproduction version of the mesh gateway has no traffic.

The mesh gateway sends traffic to the backend services, A and B, both of which send traffic to C. Service B is in the process of a new release, and has two production versions, the old version and the newly release version, with blue-green traffic shifting. Traffic into service B is randomly assigned to the blue or green node.

For service A, a preproduction version (see V.E) is running its final regression tests before entering the blue-green traffic shifting phase.

Production traffic into services A and B uses the production databases, but all other traffic use staging databases (see V.F).

The scale of this architecture at Coursera is shown in table 1, where we note the JS apps are automatically deployed to preproduction, but all other steps require at least one manual button click.

### E. Preproduction Deploys

A preproduction deploy is a deploy that may not serve production traffic. Within the CND architecture, *all* deploys are preproduction deploys, and production traffic is not directed to them. Before being released to production, the deploy must be validated with a successful run of the integration tests. After this, an engineer can start the blue-green traffic shifting process, or automatic canary analysis, converting the preproduction deploy into a production release.

Deploys that are built from the main branch are eligible for traffic shifting and release. Only appropriately approved code may be merged to the main branch; thus, we control the code which handles production traffic.

For all preproduction deploys an audit trail keeps the commit details.

When a preproduction deploy is created appropriate URLs are also created to allow developers to easily route their requests (whether from their browser or from tools like **curl** and **postman**) to the selected versions, for easy ad hoc testing and troubleshooting.

### F. Isolation with Staging Databases

To provide isolation of preproduction from production, test traffic routes via a different set of databases. Staging databases are copied on a regular cadence (e.g., daily), from production, with sensitive data excised (see VII.A and VII.B).

Using a copy-on-write approach to cloning makes this relatively fast and cheap, but even full copies are feasible. It may suffice to have partial copies of some databases. Schema migrations are included in the staging database by mirroring schema and routing changes in the schema migration code of the owning service.

The low-level libraries that access databases must check on a per request basis whether to use the production database or the staging database: we describe such database access as Staging Aware.

*1) Ids in Databases:* When creating new data, it is helpful if the ids used in the production database do not collide with ids used in the staging database. This can be achieved for example, by using UUIDs [21]. For auto increment fields we modify the offset when making the staging DB, to be much higher than the production value. We vary the starting point each day to further isolate each day's staging database from the previous day.

*2) Cache Collisions:* Architecturally, each cache over Staging Aware databases should in turn, be Staging Aware. However, given the above care with ids (which is easy), in practice almost all our caches work fine, without being Staging Aware, due to different usage patterns. This is useful when migrating to this architecture, since making the caches Staging Aware can be done at the end of an incremental process.

### G. Integration Testing

As with any CI/CD architecture, it is recommended to have a comprehensive set of integration tests that capture the business-critical user flows.

As part of the architecture, these tests can be run with routing annotations selecting specific components, which are then the components under test for regression testing.

Tests which modify databases must use the staging databases. It is also critical that these tests are parallel safe: i.e., the same test can be run concurrently both with itself and with any other test. While non-parallel safe tests may still be useful, they cannot be part of the integration test suite.

*1) Test Data:* Typically, integration tests require known test data. One approach is to have a special test database, and test system, that is kept up to date with production. In practice, maintaining version parity between all the components is often difficult.

In contrast, the CND architecture has:

- Static test data in the production database, explicitly labelled as fake, and excluded from certain product views. This is copied into the staging database as part of the normal regular process.

- And dynamic test data created, perhaps by copying static test data, during test setup.

By keeping the static test data in the production database, it automatically is subject to any schema migrations that are applied.

*2) Synthetic Tests:* The integration test suite can be used unchanged as a synthetic test suite, running regularly against the production system, with the staging databases.

## VI. Advantages & Disadvantages

We turn from describing the architecture to advocating for it. The main advantages are:

- Ease and quality of integration testing, ad hoc testing and troubleshooting with high non-functional fidelity since the system used for testing (production with a few preproduction components, and the staging databases), is extremely similar to production.
- Reduced need for special testing infrastructure, with lower maintenance cost: in the place of the conventional staging system being a full (but scaled down) copy of production, we simply have the staging databases.
- That the architecture allows incomplete implementation with pragmatic prioritization alongside other business goals.
- That a monolithic security goal (physical isolation of preproduction and production), is broken down into more precise and better motivated goals for data isolation which can be prioritized more accurately.

Disadvantages include:

- Novel failure modes.
- The CND environment fails with production.

### A. A Scalable Test Environment that Mirrors Production

As described in section II.B and [16] the primary goal of an integration testing environment is to be very like production. This is achieved exceptionally well by this architecture, a typical test activity uses one preproduction deploy with the rest being the production system. Unlike other approaches, almost all non-functional aspects of production are captured. This scales with the number of developers, independently of the total system size. Alternative approaches often clone the whole system. Such a clone can only be used by one team at a time, without explicit co-ordination.

### B. New Failure Modes

It is possible, when following this architecture to have new types of production failure relating to preproduction deploys.

In addition, traffic patterns for testing may differ from normal traffic. Much testing will be for your most frequently exercised code paths for which test traffic and production traffic will be proportionate. In addition, you will also want to test less frequently used code paths of high value. For example, Coursera typically launches zero, one or two new degrees each month, but the test system creates hundreds of new degrees each day, in the staging databases. Thus, the testing requirements put qualitatively different stresses on some production components.

In our experience, problems arising have been minor, and each specific issue is addressed like other production issues, with fixes prioritized alongside other potential activities.

### C. Preproduction goes Down with Production

A further disadvantage of making extensive use of preproduction for testing during development lifecycle is that production outages are also preproduction outages. However, the internal production service level objectives [22], [23] (we target 99.95%) means that the target uptime for the development environment is high.

### D. Pragmatic Implementation of the Architecture

The architecture acts as a guide and is intended to be partially implemented. This allows the prioritization of each feature for each component to be managed independently. While each component type is important, if there are sufficiently few engineers actively developing such components, then fully automatic support may be lower priority than other work.

In practice, we have found that staging databases and staging aware caches V.F can be rolled out one-by-one, in contrast with a traditional staging system, that tends to be all-or-nothing.

### E. Managing Risk

Cloud services involve risk. It is necessary to make changes to the running software, to meet new business objectives, yet making those changes involves occasional failure. Coursera, like many cloud companies, operates in "perpetual development" [24]. We are "continuously develop[ing] new features" rather than just maintaining the current platform.

Two key concepts are the error budget [22, chap. 3], and blast radius [25]. Given an internal service level objective (SLO) of 99.95% then that leaves 0.05% of acceptable errors, or just over 20 minutes a month of "error budget". Unlike a one-sided service level agreement, the error budget is used to "to balance service reliability with the pace of innovation" [26]. Adjusting the SLO allows you to choose a different balance between the two.

*1) Choosing to Spend the Error Budget on Development:* The point of the error budget is to find the "right balance between innovation and reliability" [22]. The CND architecture introduces new failure modes, which involve spending some of the error budget. Our experience is that the quality improvements from allowing developers to truly test their code in production early in the cycle offset this cost, and can promote increasingly rapid innovation. The production downtime associated with the new risks is, in practice, low. We also believe that the production downtime averted, because of the advantages identified throughout this paper, is significantly greater.

*2) Tight Coupling as Risk:* Almost all the many changes we make are within a single component that is loosely coupled with the rest of the system, so that the change can genuinely be independently deployed: and by using the CND architecture the change can be deployed safely.

Some changes reveal a tight coupling between supposedly independent components. One area is upgrades to the service mesh. For these, a naïve use of the CND architecture is inadequate, but other testing and quality assurance (QA) strategies must be used. Another example is near-line processing of events in a message stream. Such streams couple a producer and a consumer: to test we can use preproduction deploys of both, with a new non-production stream, used to isolate this development activity from production.

Other QA strategies include: extended review; clear monitoring and rollback plans; use of one-off custom test infrastructure; de-risking the blast radius of the change. Such high-risk changes are blocked when the error budget is depleted.

*3) Focus on Effective Security Strategies:* Understand actual threats, and have a risk mitigation approach for these.

## VII. Security Considerations

Both PII and other sensitive data should be removed from the staging databases both for regulatory conformance, and as sound business practice.

### A. Personally Identifying Information (PII)

Production databases generally have some PII in them. Appropriate use of this data is narrow including running the actual production system. It is not needed for testing which is the focus of the preproduction deploys. Since these use the staging databases, the recommended approach is to remove the PII and other sensitive data when copying the data from production to staging. True de-identification is difficult and aims to make re-identification impossible, even when the adversary has a full database dump in their possession [27], [28]. More realistically, and appropriately, using some version of HIPAA's limited data set [29] (with the addition of password fields) addresses key risks of concern (insiders snooping on relatives or friends or famous people) while being much easier. This involves erasing or randomizing certain key fields which identify the users, while leaving the other data intact.

### B. Sensitive Data

A specific threat is that employees may seek access to material non-public information to illegally trade more profitably on the stock market. The CND architecture has the following safety features:

- Each production database with sensitive data is only accessible from production deploys of one microservice.

- Preproduction deploys should only be able to access the staging database.

When creating the staging database sufficient fields are randomized to leave the resulting data valueless

### C. Security of Preproduction Requires New Mindset

Verifying that data isolation addresses security threats is different from verifying that isolation of the complete environment addresses similar threats.

Good practice is to identify the specific threats one wishes to protect against, and to make proportionate response. [30] identifies financial gain (for current employees) and sour grapes (for former employees) as two principal motivations. A further threat is that an external bad actor compromises an employee's laptop. Data isolation addresses these by preventing access to production data from preproduction, and by removing sensitive personal and financial data from the staging database. [31] says: "Risk analysis should be obviously the first step before implementing […] countermeasures, as depending on the risk profile of the organization, implementation of all countermeasures may be inappropriate and result in performance degradation and high administration costs." In our opinion, motivating the use of entirely separate environments as opposed to data isolation in order to improve security, requires a compelling risk analysis that meets this high bar. For our principal use case, of running integration tests just before release, the key threat is that changes made by the end-to-end tests will cause database degradation (e.g., by filling the databases up with test data or by having hard deletes that may incorrectly delete product data). This risk is mitigated through the staging databases.

### D. Routine Security Tasks

A further consideration is that routine security tasks, like OS upgrades, are simplified through having fewer cloud environments thereby reducing both risk and maintenance costs and toil for CI/CD teams. Such security updates to a separate staging environment often are de-prioritized and can become a vector of attack from bad actors.

## VIII. Implementation Considerations

This section gives the steps to implementing this architecture on top of a cloud application that is already cloud native (see V.A).

For all the components that are to be tested for regression by the integration testing you must first ensure that you can deploy multiple versions. Your deploy and release tools should use this capability to separate deploy from release (V.C). and switch production from one to the other, both for production traffic, and independently for test traffic. New annotations need to be added to the traffic to control the switching for these components (V.D).

Assuming the use of a service mesh, on service mesh ingress you must ensure that an overall testing flag is set if the incoming request is annotated for any preproduction component. In addition, such annotations need to be cryptographically verified on ingress.

You must establish the process of copying the staging databases (V.F), for example on a daily cadence. This copying must be integrated either with your pre-existing de-identification system, or you must create one.

You must ensure that your integration test frameworks (V.G) support setting the annotations, on a per request basis, to route through the preproduction components under test.

At this point, you are able to run regression integration tests manually.

You may need to make some caches staging aware (V.F.2).

If tests depend on near-line processing via message queues, the routing annotations need to be added to the message envelopes.

To complete the implementation, you modify the deploy and release tools to call the regression test, as an obligatory step, between deploy and release.

## IX. Conclusions and Future Work

We have shown how using preproduction deploys within the production cloud, along with the use of staging databases, and separating release from deploy, can allow integration tests to be used for regression testing immediately before production release of a cloud component.

This testing is scalable, without either the artificial bottleneck of a single staging system, or its cost.

This testing is of high quality since the system under test is the very same system that is released immediately after a successful test run.

We would like to show how other development activities can also be safely performed using such preproduction deploys. We hope, in the future, to clearly articulate: the guard rails needed to allow developers such access without compromising the integrity of production; and to explore the costs and benefits of so-doing.

## Acknowledgment

Many thanks to Luis Motta Campos, Howard Shin, Sasha Usenko, Christopher Watkins, Richard Wong and the anonymous referees for very helpful review comments.